# DARK MATTER, DARK ENERGY AND ROTATION CURVES

**C. SIVARAM** and **VENKATA MANOHARA REDDY.A***

Indian Institute of Astrophysics,
Bangalore, 560 034
India

## ABSTRACT

Models for DM are invoked with various density distributions, to account for flat rotation curves of galaxies. The effect of DE on these profiles for large galaxies and clusters are also studied. The rotation curves are shown to dip at a particular distance, when the effect of DE became significant.

## Introduction

Observations imply that there is a considerable amount of dark matter or unseen matter on astronomical scales ranging from clusters of galaxies to individual galaxies themselves. For instance the masses of galaxies in a cluster as estimated from the virial theorem to account for their observed velocity dispersion ($v^2$) giving the dynamical mass, $M_d \sim (v^2)R/G$ turns out to be at least a factor of ten higher than what one would except from the luminosity(Faber and Gallagher 1979). Even groups of galaxies seem to have inadequate luminous mass by a similar factor. To account for their dynamical dispersion, the proportion of unseen non luminous mass should increase with increasing scales. Again studies of the dynamics and structure of large spiral galaxies suggest that a universal feature of all the rotation curves is that at large galacto centric distances they are either flat or slowly rising, there being no large spiral galaxy whose rotation curve falls (Rubin et al. 1982). The rotational velocities for a point mass (keplerian) are given by $v^2$ being proportional to $GM_r/r$, $M_r$ being the mass contained within a radius r. These observations of flat , v = constant, rotation curves imply $M_r$ increasing linearly with r indicating the presence of much unseen dark matter up to large distances from the centre of the spiral galaxies. The progressive increase in the dynamical mass with radius is a characteristic feature of all these galaxies, i.e. individual galaxies are surrounded by massive dark halos, which have as much as ten times the mass of the visible matter. It is now known that x-ray emitting hot gas (e.g. from clusters and galactic coronae) would account for only a small fraction of the required missing mass. Other propositions for DM ranging from black holes to very low mass stars have met with various difficulties, So finally the presence of dark matter in halos and beyond halos (in clusters) imply a large ratio of dynamical mass to luminous mass. This non baryonic mass is present for large distances from the galaxy. The orbital velocity remains constant at larger distance from the galactic core.

The galaxy as has long been suspected has at its centre a massive black hole, with estimated mass of around 3million suns. If the galaxy was held together by the attraction of that mass, and the motion around it was circular, if `G' is the constant of gravitation, the balance of forces on any mass m in the galaxy is

$$\frac{GMm}{r^2} = \frac{mv^2}{r} \qquad (1)$$

Giving

$$v^2 = \frac{GM}{r} \qquad (2)$$

The only source of attracting gravity was a central black hole. Its mass would provide M, and velocity would diminish as $1/\sqrt{r}$, as it does in the solar system. However, galaxies contain a lot more mass than their central bodies. The Milky Way galaxy itself contains about a trillion solar masses.

**The effect of Dark Matter (DM) on rotation curves**:

Thus the effective value of M increases with distance, and the rotation velocity v in the denser parts of the galaxy may falls off less steeply than like 1/r. However, beyond densest part, v should drop off, and this fall off should be close to 1/r. But in practice the velocity of the objects beyond the halos become constant. The matter implies still present but its not radiating. One can give some models.

For example consider a gas of collision less particles held only by their mutual gravity (sivaram, 1987). Velocity of the particle is v and mass is $m_d$(DM particle). Then average kinetic energy of the DM particles can be written as,

$$k = \tfrac{1}{2} m_d v^2 \qquad (3)$$

And total mass with in the radius r can be written as

$$M_{(r)} = \frac{v^2 r}{G} = \frac{2kr}{m_d G} \qquad (4)$$

$$\frac{dM_{(r)}}{dr} = \frac{2k}{m_d G} \qquad (5)$$

For hydrostatic equilibrium (jeans, 1925)

$$\frac{dM_{(r)}}{dr} = 4\Pi r^2 \rho_{(r)} \qquad (6)$$

From the above equations we can write density as a function of, r

$$\rho_{(r)} = \frac{2k}{4\Pi r^2 m_d G} \qquad (7)$$

This equation implies that $\rho_{halo} \propto 1/r^2$

From this one can calculate the number density ($n_d = \rho/m_d$). Let us take velocity 300 km/sec, mass of the particles $m_d$ is ~ 10ev and the radial distance 10kpc, than the particles must have number density ~ $10^8$ m$^{-3}$.

If we take $\rho_{halo} \propto 1/r$ we face some problem when comparing this model with D.E. We will come back to this later in this section. When $\rho_{halo} \propto 1/r^3$ the mass term becomes constant so that the velocity falls quickly after the galaxy edge. So the suitable model with which one can fit flat curves is one for $\rho_{halo} \propto 1/r^2$ where $\rho_{halo}$ is the dark matter density in halo and beyond halo as a function of r. As stressed in Sivaram (1985, 1987), the DM particle cannot be degenerate. For a mass distribution dominated by degenerate particle, the mass—radius relation is of the form $M \sim R^{-3}$, and this would by no means imply a flat rotation curve. For DM particle of 10ev mass and density $10^8$ m$^{-3}$, the degeneracy energy is far smaller than their thermal or kinetic energies. Phase space constraints are also discussed in Sivaram (1987) and also is the presence of a cosmological constant (Sivaram, 1985).

Now consider,

$$\rho_{halo} = \frac{\rho_o R^2}{R^2 + r^2} \tag{8}$$

Where $\rho_o$ is the core density of the galaxy, R is the core radius, r is the halo radius.

Then the mass with in the radius r is can be written as

$$M_{(r)} = \left[\frac{4\Pi\rho_0 R^3}{3}\left(-2 - 3\tan^{-1}\left(\frac{r-R}{R}\right)\right) + 4\Pi r\rho_0 R^2\right] \tag{9}$$

Then the radial velocity term for any point mass in the halo and beyond halo can be written as

$$v = \left[\frac{4\Pi G\rho_0 R^3}{3r}\left(-2 - 3\tan^{-1}\left(\frac{r-R}{R}\right)\right) + 4\Pi G\rho_0 R^2\right]^{(1/2)} \tag{10}$$

If r>>R we can re write the above equation as

$$v = \left[\frac{4\Pi G\rho_0 R^2}{r}\left(r - \tfrac{2}{3}R\right)\right]^{(1/2)} \tag{11}$$

From the above equation one can calculate the rotational velocity of any point mass if it is far away from the galaxy centre. Equation (10) also implies the same thing. From equation (10) when we calculate the velocity versus distance the curve takes a small dip at the edge of the galaxy, and then rises and remains constant for very large r. This is shown in the figure below.

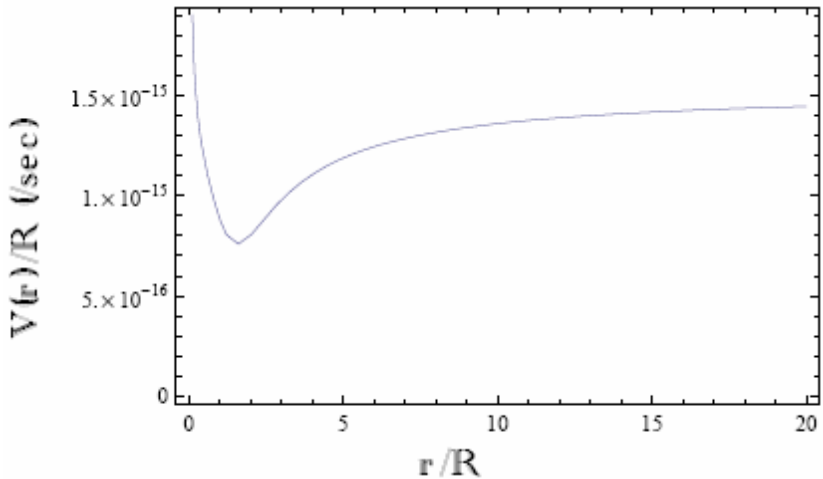

The velocity does not dip as r increases in the presence of dark matter if $\rho$ continues to fall as $1/r^2$. Now it's the time to think about dark energy.

**What about Dark Energy (DE):**

Current cosmological observations imply a universe dominated by dark energy with its associated negative repulsive force (Dodelson and Knox, 2000, Ostriker and Steinhardt, 2001). This negative repulsive pressure of Dark Energy (DE) may disrupt all bound structures (Sivaram, 1999, Caldwell, 2000.). Dark energy is characterized by negative pressure $p = \omega \rho c^2$, $\omega = -1$, corresponding to the cosmological constant introduced by Einstein. In GR, pressure also contributes to gravity. So $p = -\rho c^2$, implies a, positive acceleration, and not deceleration as in the case of attractive gravity, but repulsion. At present, DE constitutes 0.7 of the energy in the universe as implied by supernovae (SN) and WMAP observations. We have the scalar potential from a metric of the Schwarzschild--desitter form (Misner, Thorne, Wheeler, 1973).

$$\Phi = 1 - \frac{2GM}{rC^2} - \Lambda \frac{r^2}{3} \qquad (12)$$

The force implied by the above equation can be equated with the rotational centrifugal force. We get the velocity term, as

$$v = \left[ \frac{GM}{r} - \Lambda \frac{r^2 C^2}{3} \right]^{(1/2)} \qquad (13)$$

By combining this velocity term with that of the previous equations 10, 11 we get the modified equations for the velocity in the presence of DE .This turns out to be;

$$v = \left[ \frac{4\Pi G\rho_0 R^3}{3r}\left(-2 - 3\tan^{-1}\left(\frac{r-R}{R}\right)\right) + 4\Pi G\rho_0 R^2 - \frac{\Lambda C^2 r^2}{3} \right]^{(1/2)} \qquad (14)$$

And,

$$v = \left[ \frac{4\Pi G \rho_0 R^2}{r}\left(r - \tfrac{2}{3}R\right) - \frac{\Lambda C^2 r^2}{3} \right]^{(1/2)}, \text{ respectively} \qquad (15)$$

The above two equations gives the velocity of the point masses beyond the halos in the presence of the dark matter and dark energy. By plotting the curve for both with and with out the $\Lambda$ term applying equations 14, 15 and 10,11 respectively. The curves imply in the first case absence of dark energy term. The curve never dips it will monotonically remain constant for very long distances. In both these cases it means the dark matter increases with r. And in the initial stages, and the moment it reaches the galaxy edge, it will take a small dip and then it will rise. Then the density of the dark matter follows the $1/r^2$ profile the velocity remains constant. If we will apply the dark energy term as shown in equations (14) and (15) the velocity decreases at this point when both the terms are comparable. On small scales the first term in the above equations dominates and the second term is not important for the rotational curves. This is shown in the following figures.

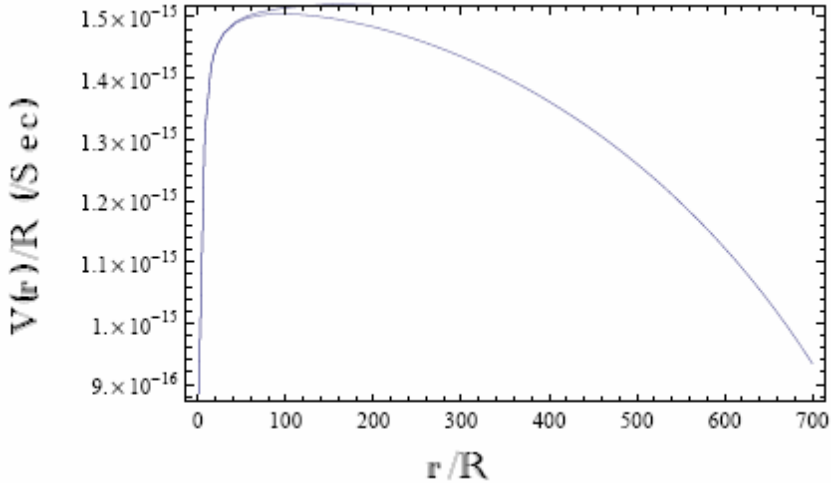

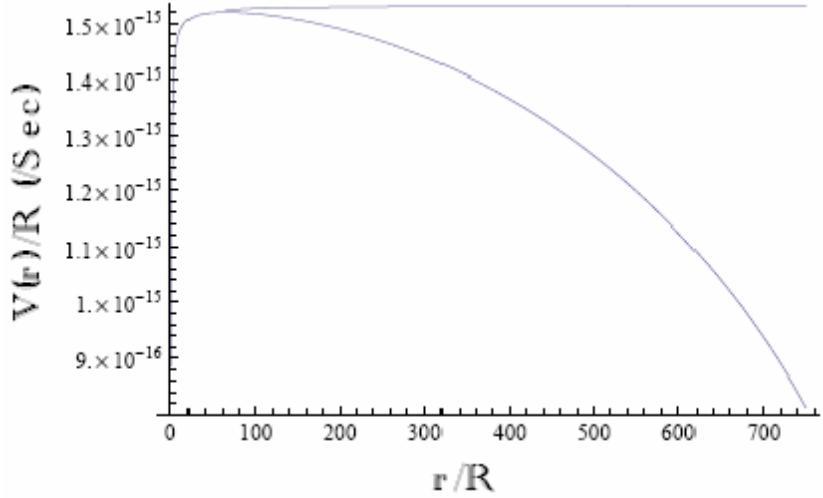

In the above figures the constant velocity curve shows orbital velocity of a point mass with respect to distance. The curved line is due to presence of dark energy term at long distances, where it slowly comes down. General relativity suggests that the cosmological constant if present exerts a repulsive gravity force, on large scales. So by considering this term we can generalize the above equation for the distance up to where its effects can be felt.

From the above equations we can estimate the distance at which the DE starts affecting the rotational velocity of any massive object, in the ideal situation where there is no other object near by it. In other words, at what distance the gravitational effect of any object will begin to be affected by the back ground DE? Equation (13) indicates that net force would vanish, i.e.

$$m \times \nabla \Phi = m \times \left[ \frac{2GM}{r^2 C^2} - \frac{2\Lambda r}{3} \right] = 0 \tag{16}$$

When the distance beyond the mass M is,

$$r = \left[ \frac{3GM}{\Lambda C^2} \right]^{(1/3)} \tag{17}$$

This formula gives the distance at which the $\Lambda$ term starts to dominate. $\Lambda$ cosmological constant has a value given by $10^{-56} cm^{-2}$ (as implied by current observations). Now let us calculate r for super clusters of galaxies with a mass of $10^{18}$ solar masses this gives,

$$r = 3.5 \times 10^{24} \, m$$

For a large cluster of galaxies with a mass of $10^{16}$ solar masse this gives,
$$r = 7.6 \times 10^{23} \text{ m}$$
For a mass of $10^{12}$ solar masses equation r will be
$$r = 3.5 \times 10^{22} \text{ m}$$
This limit is applicable only in the situation where other objects are absent .One can test this for a cluster of galaxies where the effect is more. The distance exceeds the value given by the equation (17). The negative pressure implied by $\Lambda$ counteracts the gravitational attraction of that particular mass.

Even in selecting the density of dark matter we can not use the case $\rho_{halo} \propto 1/r$ because in that case the velocity term in the equation (13) rises as $r^2$ .If we apply this density profile even the first term rise. This doesn't make sense. If we take $\rho_{halo} \propto 1/r^3$, the mass is constant, so that the orbital velocity decreases rapidly as soon the baryonic disk is crossed. After this distance one can use the density model for DM halos where $\rho_{halo} \propto 1/r^2$. And this model is compatible with the presence of dark energy at long distance scales.

## What if ω is not equal to -1 in the equation of state for DE :

Now the equation of state for dark energy is written as (Peacock, 2000.)
$$p = \omega \rho c^2, \qquad (18)$$
Where ω is a number equal to -1 $\rho$ is density of the dark energy, c is the velocity of light. If ω is of different value, then a generalized metric can be written as

$$g = g_0 \left(1 - \frac{2GM}{r^2} - \frac{A}{r^{3\omega+1}}\right) \qquad (19)$$

Here A is a constant. If we consider ω = -1, A= $\Lambda$/3. If ω=-1/3, A = 0 then in the equation (13), $v^2$ becomes $GM/r^2$, so this term is general. If we take ω is positive then the velocity in equation (13) become imaginary. So from this one can conclude that ω value must be<-1/3.

So finally in the equation of state one can use the value for ω as mentioned above for the generalized case.

## Concluding Remarks:

Several observations imply that the rotational curves of the galaxies for long distances are flat. It indicates the presence of DM or unseen matter.

Considering suitable models for these DM halos, one can plot the flat rotation curves. If we consider the effects of cosmological constant on large scales, the flat curves take a dip at very large distances. From this we can get the formula for the distance beyond which the dark energy dominates. By applying different values for ω in the generalized metric, we conclude that the ω value should be always< -1/3.

**[*Acknowledgement:** One of the authors (VENKATA MANOHARA REDDY.A) is grateful to Indian Institute of Astrophysics for providing the requisite facilities.**]**